\documentclass[aps,prl,superscriptaddress,twocolumn,longbibliography]{revtex4-2}
\usepackage{bbm}
\usepackage{graphicx}
\usepackage{dcolumn}
\usepackage{bm}
\usepackage{subfigure}
\usepackage{amsmath}
\usepackage{amssymb}
\usepackage{feynmf}
\usepackage{hyperref}
\usepackage{color}

\usepackage{attachfile}

\usepackage{times}

\begin{document}


\title{Magneto-Transport Properties of Kagome Magnet TmMn$_6$Sn$_6$}

\author{Bin Wang}
\affiliation{Center for Neutron Science and Technology, Guangdong Provincial Key Laboratory of Magnetoelectric Physics and Devices, School of Physics, Sun Yat-Sen University, Guangzhou, 510275, China}

\author{Enkui Yi}
\affiliation{Center for Neutron Science and Technology, Guangdong Provincial Key Laboratory of Magnetoelectric Physics and Devices, School of Physics, Sun Yat-Sen University, Guangzhou, 510275, China}

\author{Leyi Li}
\affiliation{Center for Neutron Science and Technology, Guangdong Provincial Key Laboratory of Magnetoelectric Physics and Devices, School
of Physics, Sun Yat-Sen University, Guangzhou, 510275, China}

\author{Jianwei Qin}
\affiliation{Center for Neutron Science and Technology, Guangdong Provincial Key Laboratory of Magnetoelectric Physics and Devices, School of Physics, Sun Yat-Sen University, Guangzhou, 510275, China}

\author{Bing-Feng Hu}
\affiliation{Key Laboratory of Neutron Physics, Institute of Nuclear Physics and Chemistry, China Academy of Engineering Physics, Mianyang, 621999, China.}

\author{Bing Shen}
\email{Corresponding author: shenbingdy@mail.sysu.edu.cn}
\affiliation{Center for Neutron Science and Technology, Guangdong Provincial Key Laboratory of Magnetoelectric Physics and Devices, School of Physics, Sun Yat-Sen University, Guangzhou, 510275, China}
\affiliation{State Key Laboratory of Optoelectronic Materials and Technologies, Sun Yat-Sen University, Guangzhou, Guangdong 510275, China}

\author{Meng Wang}
\email{Corresponding author:wangmeng5@mail.sysu.edu.cn}
\affiliation{Center for Neutron Science and Technology, Guangdong Provincial Key Laboratory of Magnetoelectric Physics and Devices, School of Physics, Sun Yat-Sen University, Guangzhou, 510275, China}

\begin{abstract}
Kagome magnet usually hosts nontrivial electronic or magnetic states drawing great interests in condensed matter physics. In this paper, we report a systematic study on transport properties of kagome magnet TmMn$_6$Sn$_6$. The prominent topological Hall effect (THE) has been observed in a wide temperature region spanning over several magnetic phases and exhibits strong temperature and field dependence. This novel phenomenon due to non-zero spin chirality indicates possible appearance of nontrival magnetic states accompanying with strong fluctuations. The planar applied field drives planar Hall effect(PHE) and anistropic magnetoresisitivity(PAMR) exhibiting sharp
disconnections in angular dependent planar resistivity violating the empirical law. By using an effective field, we identify a magnetic transition separating the PAMR into two groups belonging to various magnetic states. We extended the empirical formula to scale the field and temperature dependent planar magnetoresistivity and provide the understandings for planar transport behaviors with the crossover between various magnetic states. Our results shed lights on the novel transport effects in presence of multiple nontrivial magnetic states for the kagome lattice with complicated magnetic structures.
\end{abstract}
\date{\today}
\maketitle

\section{Introduction}
Kagome magnets have emerged as an important platform to study electronic correlations and nontrivial topology\cite{J. X. Yin, L. Ye, E. Liu, M. Kang, Y. Kasahara, E. Tang}. The unique crystal structure, made of corner-sharing triangles, naturally has relativistic band crossings at the Brillouin zone corners hosting nontrivial topological electronic Dirac Fermion and a dispersionless flat band\cite{K. Ohgushi}. With inclusion of spin orbital coupling and magnetism, the system can host various nontrivial topological electronic or magnetic states such as  magnetic Weyl Fermions, quantized anomalous Hall states, and non-zero spin texture etc \cite{J. X. Yin, X. Xu}. Fabricating various magnetic structures in kagome compounds would effectively engineer exotic states with novel phenomena to study the interplay between electron and magnetism in condensed matter physics.

Recently, a rare-earth family of kagome magnets ReMn$_6$Sn$_6$ (Re is the rare earth elements) with a layered hexagonal structure has drawn great interest due to the rich magnetic states and  various nontrivial topological band structures\cite{N. J. Ghimire, M. Li, Q. Wang, W. Ma, L. Gao, G. Venturini, B. Malaman1, D. M. Clatterbuck}.  These materials usually consist of two sub magnetic lattices from different layers: (1) the magnetic kagome lattice made by Mn ions. (2) another magnetic lattice made by rare earth ions. The intra- or inter- layer magnetic interactions drive various magnetic structures and complicated magnetic phase diagrams. TmMn$_6$Sn$_6$ is typical one of these materials\cite{B. Malaman, Y. Sawai, L. K. Perry, D. M. Clatterbuck}. It forms a hexagonal P6/mmm structure (a = 5.514 \AA,  and c =8.994 \AA) consisting of kagome planes Mn$_3$Sn separated by two inequivalent Sn$_3$ and Sn$_2$Tm layers as shown in Fig. 1(a). The temperature dependent resistivity reveals TmMn$_6$Sn$_6$ is a good metal with small anisotropy. Observed sharp kinks around 325 K in temperature dependent magnetization ($M(T)$) curves reveal a magnetic transition shown in Fig. 1(d). Below this transition, the  collinear-antiferromagnetic structure transits to a  heli-magnetic structure  which is revealed by neutron scattering measurements and predicted to host chiral spin texture such as skymions\cite{S. W. Cheong, C. Lefevre, N. Nagaosa1}. Compared the light rare earth elements such as Y, Tm is a heavy rare earth element with $4f$ electron and strong spin-orbital coupling which may lead to stronger magnetic interactions hosting more complicated magnetic states.

In this paper, we systematically studied the magneto-transport properties of TmMn$_6$Sn$_6$. The prominent topological Hall effect are observed with an applied field in the $ab$ plane suggesting the emergence of exotic magnetic states with strong fluctuations. In presence of magnetic transitions, the planer Hall effect (PHE) and anisotropic magnetoresitivity (PAMR) do not obey the empirical law. Especial for PAMR,  the abrupt disconnections due to magnetic moment flops separate planar longitudinal resistivity $\rho_{xx}^P$ into different groups belonging to various magnetic states. To describe these novel transport behavior, we scaled $\rho_{xx}^P$ by defined an effective field $\mu_0H^e$ to acquire the critical field. By modifying the empirical law we scaled angular dependent $\rho_{xx}^P$ at various fields and provide the understanding for these planar transport behaviors spanning over various magnetic states.

\begin{figure}
  \centering
  \includegraphics[width=3.2in]{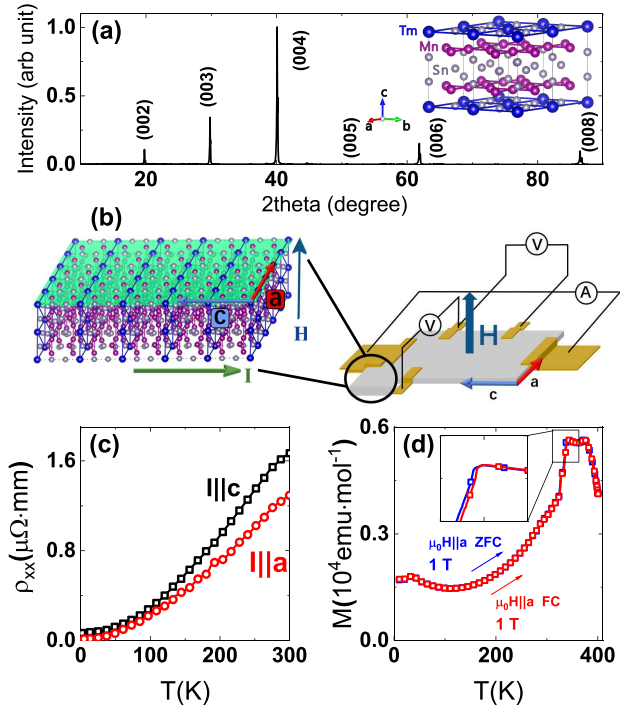}
  \caption{(a)X-ray diffraction pattern of a single crystalline TmMn$_6$Sn$_6$ with the corresponding Miller indices (00l) in the parentheses. Inset: crystal structure of TmMn$_6$Sn$_6$. (b)Schematic for our electronic measurement device of TmMn$_6$Sn$_6$. (c) Temperature dependent longitudinal resistivity of TmMn$_6$Sn$_6$ with the current along the $c$ axis(black line) and $a$ axis(red line) (d) Temperature dependent magnetization of TmMn$_6$Sn$_6$ with $\mu_0 H$=1 T of $\mu_0 H//a$ zero field cooling process(ZFC)(blue line),$\mu_0 H//a$ field cooling(FC)(red line),they almost overlap.Inset: partial enlargement of the curve.
 }
  \label{fig:Fig1}
\end{figure}
\section{Experimental Detail}
Single crystals of TmMn$_6$Sn$_6$  were grown via the self-flux method. Tm, Mn, and Sn metals were mixed and sealed inside an evacuated quartz tube. After that, the mixture was first heated to 1100$^{\circ}$C and then slowly cooled to 550 $^{\circ}$C where flux was removed by using a centrifuge. Large plates of single crystals are obtained with a typical size of 5 $\times$ 5 $\times$ 1 mm$^3$. The crystal structure and elemental composition were confirmed by X-ray diffraction (XRD) measurements and Energy-dispersive X-ray spectroscopy (EDS). The [00$l$] peaks were observed in XRD patterns indicating the high quality of our crystals shown in Fig. 1(a). The selected single crystals were shaped into a rectangular slice for electric and magnetic transport measurements. The six gold contacts were make on the $ac$ plane with the current along the $c$ axis of crystals as shown in Fig. 1(b). Electrical transport measurements were performed in the physical properties measurement system (PPMS Dynacool, Quantum Design). Magnetization measurements were performed in the vibrating sample magnetometer (VSM) module of the PPMS.

\section{Results and discussion}
\subsection{Topological Hall Effect}
  \begin{figure}
  \centering
  \includegraphics[width=3.4in]{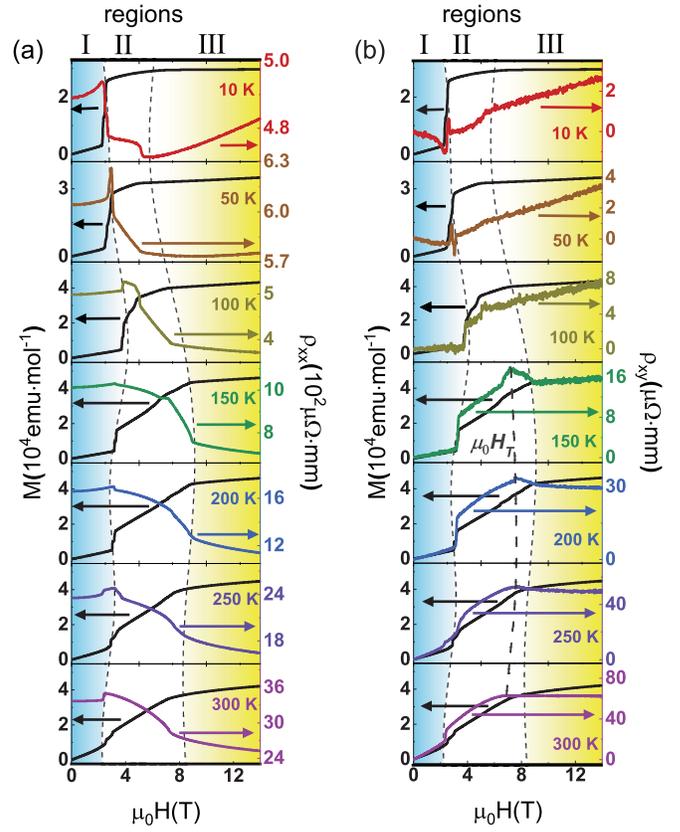}
  \caption{(a) The field dependent Magnetization $M(\mu_0 H)$ and longitudinal resistivity $\rho_{xx}(\mu_0 H)$ at 10 K, 50 K, 100 K, 150 K, 200 K, 250 K, and 300 K with $\mu_0 H\bot ac$. (b) The field dependent magnetization $M(\mu_0 H)$ and Hall resistivity $\rho_{xy}(\mu_0 H)$ at the same temperature as shown in (a). The curves can be roughly divided into regions I,II and III based on different slopes of $M(\mu_0 H)$. Distinct peaks in Hall resistivity $\rho_{xy}(\mu_0 H)$ can be observed in the characteristic field $\mu_0 H_T$.
  }
  \label{fig:Fig2}
\end{figure}
The field dependent magnetization $M(\mu_0 H)$, longitudinal and Hall resistivity ($\rho_{xx}(\mu_0 H)$ and $\rho_{xy}(\mu_0 H)$) are shown in Fig. 2  with the applied field perpendicular to the $ac$ plane ($\mu_0 H \bot ab$). $M(\mu_0 H)$ curves exhibit several obvious kinks and non-monotonic field dependence suggesting the emergence of multiple magnetic phase transitions in consistent with former reports \cite{D. M. Clatterbuck}. Based on different slopes of $M(\mu_0 H)$, two critical fields $\mu_0 H_{c1}$ and $\mu_0 H_{s}$ are identified dividing the magnetic phase diagram roughly into three regions (I, II, and III). In fact around $\mu_0 H_{c1}$ multiple transitions are observed. But since these transitions are very close, we denote the metamagnetic transition field by a single variable $\mu_0 H_{c1}$ in this paper. With decreasing the temperature $T$ from 300 K to 150 K, $\mu_0 H_{c1}$ and $\mu_0 H_{s}$ shifts to higher fields. Below 100 K with further decreasing $T$, $\mu_0 H_{c1}$ and $\mu_0 H_{s}$ shift to lower fields gradually.
In region I ($\mu_0 H\le\mu_0 H_{c1}$ ), $\rho _{xx}(\mu_0 H)$ exhibits positive field response. With increasing $\mu_0 H $, $\rho _{xx}(\mu_0 H)$ exhibits negative magnetoresistivity in regions II and III with a slope changing around $\mu_0 H_{s}$ as shown in Fig. 2(a). Correspondingly, $\rho _{xy}(\mu_0 H)$ exhibits different field dependence in various regions with the slope changing at corresponding critical fields. Especially it is noticed that in region II, a prominent peak is observed at a characteristic field $\mu_0 H_T$. This anomaly becomes unapparent with decreasing $T$ and disappears around 100 K.

\begin{figure}
  \centering
  \includegraphics[width=3.4in]{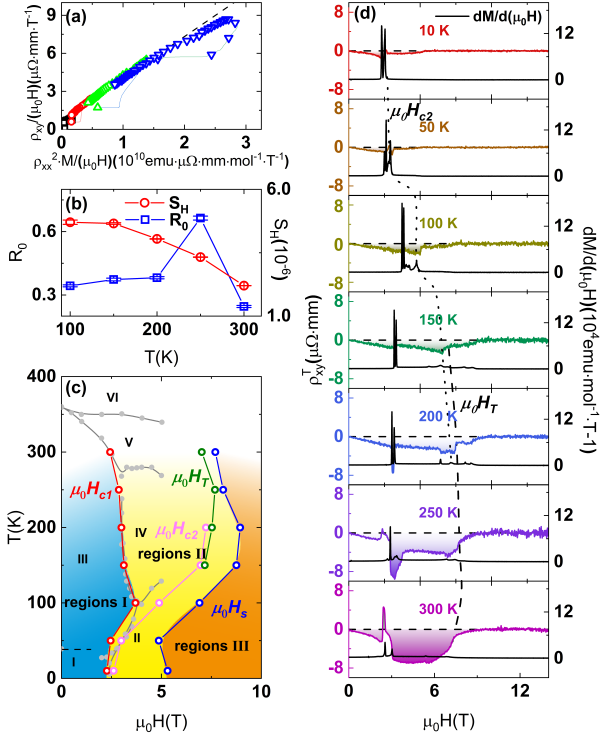}
  \caption{(a) The scaling of $\rho_{xy}(\mu_0H)/\mu_0H$ curves as a function of $\rho_{xx}(\mu_0H)^2M(\mu_0H)/\mu_0H$ at 100 K, 150 K, 200 K, 250 K. The black dash lines are the fitting curve by a linear function.  (b) The temperature dependent Hall coefficient $R_0(T)$ and intrinsic anomalous Hall coefficient $S_H(T)$. (c) The magnetic phase diagram of TmMn$_6$Sn$_6$. The data marked by gray dots or lines, and the magnetic states I, II, III, IV, V, and VI are acquired from previous results\cite{D. M. Clatterbuck}. The $\mu_0 H_{c1}$ is marked by red line agreeable with former results while the $\mu_0 H_{s}$ is marked by blue line. These two characterized field lines divide the whole phase diagram into regions I, II and III. In region II, more features are identified.$\mu_0 H_T$ (olive line) is THE characteristic peak associated with THE.  $\mu_0 H_{c2}$(magenta line) is acquired from the most obvious peak of the differential curve of d$M$/d$(\mu_0 H)$ in region II. (d) Field dependent differential d$M$/d$(\mu_0 H)$ and $\rho_{xy}^T(\mu_0 H)$ at 10 K, 50 K, 100 K, 150 K, 200 K, 250 K, 300 K.}
  \label{fig:Fig3}
\end{figure}
In a magnetic system, the Hall effect generally origins from two contributions:(1) normal Hall effect (NHE) due to the Lorentz force and (2) anomalous Hall effect (AHE) due to magnetization or Berry curvature. For TmMn$_6$Sn$_6$, the discrepancy between $\rho_{xy}$ and $M$ around $H_T$ suggests an additional contribution from topological Hall effect (THE) besides NHE and AHE. Thus, the total Hall resistivity can be expressed as:
\begin{equation}
 \rho_{xy}=\rho_{xy}^N+\rho_{xy}^A+\rho_{xy}^T=R_0 \mu_0H+S_H\rho_{xx}^2M+\rho_{xy}^T
\end{equation}
where $R_0$ is the Hall coefficient, $S_H$ is a constant for the intrinsic anomalous Hall conductivity($\sigma_{xy}^A\sim\rho_{xy}^A/\rho_{xx}^2$), which is linearly proportional to $M$, and $\rho_{xy}^N$, $\rho_{xy}^A$, and $\rho_{xy}^T$ are the normal Hall resistivity, anomalous Hall resistivity and topological Hall resistivity respectively\cite{P. K. Rout, N. Nagaosa}. According to this formula, the curves of $\rho_{xy}(\mu_0H)/\mu_0H$ vs $\rho_{xx}(\mu_0H)^2M(\mu_0H)/\mu_0H$ at various temperatures are scaled to separate various Hall contributions for TmMn$_6$Sn$_6$ shown in Fig. 3(a)\cite{Q. Wang}. By fitting the linear parts where the $\rho_{xx}^T$ term disappears $R_0(T)$ and $S_H(T)$ are acquired shown in Fig. 3(b). The $\rho_{xy}^N$ and $\rho_{xy}^A$ terms are calculated and subtracted from the total Hall resistivity to obtain $\rho_{xy}^T$.

The temperature dependence of $\rho_{xy}^T$ are presented in Fig. 3(d). The prominent THE signal is observed in high-temperature region and shrinks with decreasing the $T$. Below 100 K, the $\rho_{xy}^T$ becomes invisible. Usually THE is considered to origin from the movement of skymions in noncentrosymmertric non-colinear magnets hosting the nonzero scalar spin chirality. Recently fluctuation-driven mechanism are proposed to describe THE in centrosymmertric system relating to a special nontrivial magnetic phase\cite{N. J. Ghimire, Q. Wang}.To further understand the THE in TmMn$_6$Sn$_6$, we first analyze its magnetic phase diagram.  In our measurements, three regions are identified shown in Fig. 3(c). According to former reports,  with $\mu_0 H < \mu_0 H_{c1}$ the revealed distorted spiral (DS) orders dominate the magnetic structure in region I above 50 K\cite{N. J. Ghimire}. With $\mu_0 H > \mu_0 H_{s}$, the saturate magnetization in $M(\mu_0H)$ curves suggests a force ferromagnetism (FF) state in region III. In region II, multiple possible magnetic transitions are suggested in region II by the observed disconnections in differential field dependent magnetization d$M$/d$(\mu_0H)$ and peaks in $\rho_{xy}^T$ for TmMn$_6$Sn$_6$. In its analogue YMn$_6$Sn$_6$, transverse conical spiral (TCS) and fan-like (FL) states are observed in this region and TCS state is considered directly to relate to THE   \cite{N. J. Ghimire}. In TmMn$_6$Sn$_6$, the suggested magnetic states are more complicated than those in YMn$_6$Sn$_6$. In addition, the observed THE for TmMn$_6$Sn$_6$ probably spans over several magnetic region exhibiting strong $\mu_0H$ response in contrast to that only observed in the TCS state for YMn$_6$Sn$_6$. Thus, besides by large thermal fluctuations (such as observed in YMn$_6$Sn$_6$)\cite{A. Neubauer},  non-zero chirality leading to large THE may be also driven by appearance of possible complicated magnetic phases such as nontrival topological magnetic states. Further studies for magnetic phase diagram by neutron scattering are necessary and expected to reveal the microscopic nature for THE in TmMn$_6$Sn$_6$.

\begin{figure}
  \centering
  \includegraphics[width=3.4in]{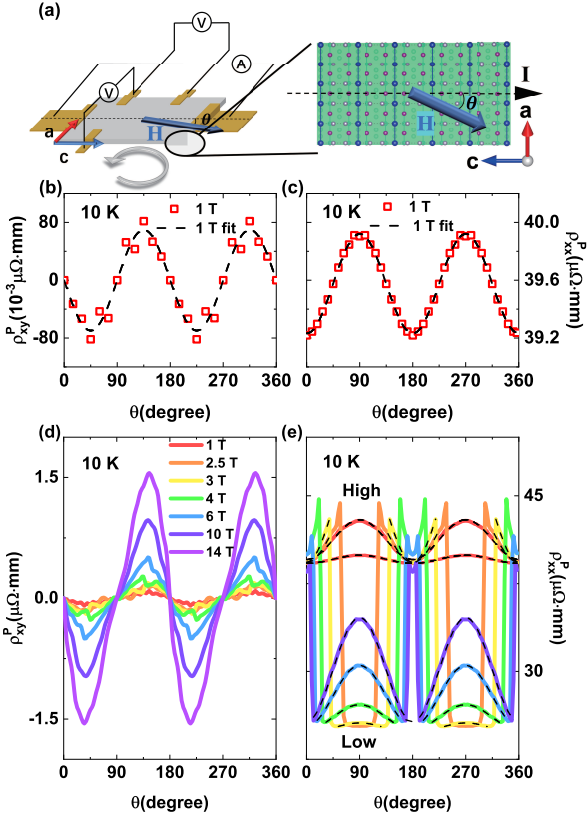}
  \caption{(a) The configuration for planar transport measurements of single crystal of TmMn$_6$Sn$_6$.The gray circular arrow indicates the rotating direction of the applied field. The current is along $c$ axis, and $\theta$ represents the angle between the applied field and current. (b) and (c) Angular dependent planar Hall resistivity $\rho_{xy}^P(\theta)$ and planar longitudinal resistivity $\rho_{xx}^P(\theta)$ at 10 K with $\mu_0H = 1 T$ respectively. The black dash lines are the fitting curves by formulas (2). (d) and (e) $\rho_{xy}^P(\theta)$  and $\rho_{xx}^P(\theta)$ at 10 K with $\mu_0 H$ = 1 T, 2.5 T, 3 T, 4 T, 6 T, 10 T, 14 T respectively.  $\rho_{xx}^P(\theta)$ is separated into ‘high’ and ‘low’ parts fitted by formula (4), marked by black dash lines.}
  \label{fig:Fig4}
\end{figure}

\subsection{Planar Hall effect and Anisotropic Magnetoresitivity }
To investigate the planar properties of TmMn$_6$Sn$_6$, the applied field $\mu_0H$ is rotated within $ac$ plane of the crystal as shown in Fig. 4(a). $\theta$ is defined as the angle between the current along the $c$ axis of the crystal (When $\mu_0H // a$, $\theta$=90$^\circ$). PHE, usually accompanied with PAMR, is a unique transport phenomenon driven by an in-plane magnetic-field-induced rotation of the principal axes of the resistivity tensor\cite{K. M. Koch, C. Goldberg} which can detect interplay of chirality, orbit, and spin for a quantum materials.  Usually, the PHE and PAMR can be described by empirical expressions as:
\begin{eqnarray}
\rho_{xy}^P&=&\Delta \rho \sin \theta \cos \theta,\\
\rho_{xx}^P&=&\rho_{\perp}^P -\Delta \rho \cos ^2 \theta,
\end{eqnarray}
where $\rho_{xy}^P$ represents the in-plane Hall resistivity that directly shows the PHE, $\rho_{xx}^P$ is the PAMR, and $\Delta \rho=\rho_{\perp}^P-\rho_{\|}^P$ is the resistivity anisotropy (called chiral resistivity in topological materials) with  $\rho_{\perp}^P$ and $\rho_{\|}^P$ representing the resistivity with the $\mu_0 H$ perpendicular (90$^{\circ}$) and parallel (0$^{\circ}$) to the current respectively\cite{K. M. Koch}. At 10 K with $\mu_0 H$=1 T, PHE and PAMR follow these empirical laws shown in Fig. 4(b) and (c). The angular dependent  $\rho_{xy}^P (\theta)$ and $\rho_{xx}^P (\theta)$ exhibit two-fold oscillations with a 45$^{\circ}$-angle shift during a whole rotating period (from 0$^{\circ}$ to 360$^{\circ}$). With increasing $\mu_0 H$, the amplitudes of oscillatory $\rho_{xy}^P (\theta)$ increase monotonously shown in Fig. 4(d) and angular dependence of $\rho_{xy}^P$ deviates from the empirical relation. Correspondingly, sudden jumps are observed in $\rho_{xx}^P (\theta)$ curves separating the data into two groups ('High' and 'Low' parts). These jumps exhibit strong field and angular dependence. The novel behaviors for PHE and PAMR are barely observed and studied before which are considered to relate to magnetic transitions for TmMn$_6$Sn$_6$.

 \begin{figure}
  \centering
  \includegraphics[width=3.4in]{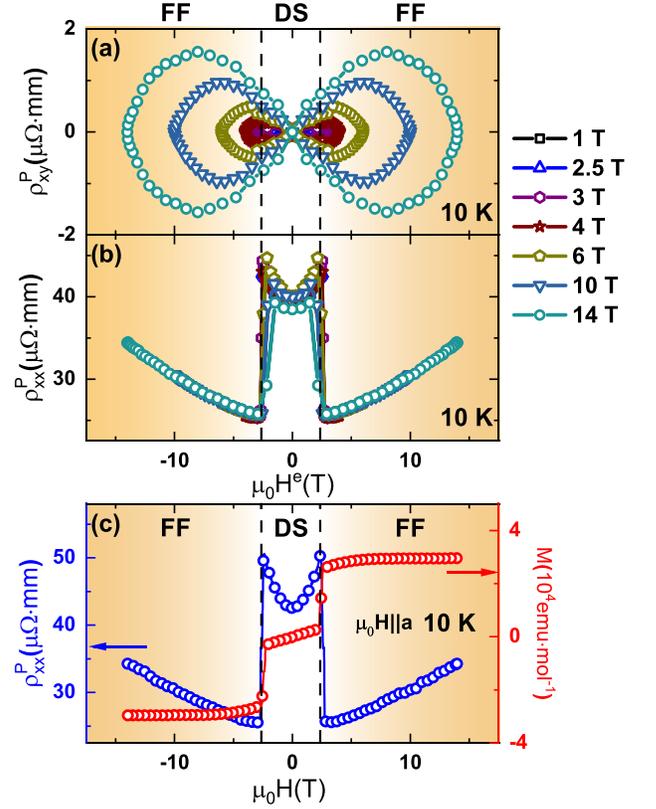}
  \caption{(a) and (b) $\rho_{xy}^P$ and  $\rho_{xx}^P$ as functions of the effective field $\mu_0H^e=\mu_0H sin\theta$ at 10 K with $\mu_0 H$ = 1 T, 2.5 T, 3 T, 4 T, 6 T, 10 T, 14 T respectively. (c) The field dependent Magnetization $M(\mu_0 H)$ and the longitudinal resistivity $\rho_{xx}^P(\mu_0 H)$ vs  $\mu_0H^e=\mu_0H sin\theta$ at 10 K with $\mu_0 H // c$. The black dash lines represent the effective critical field of $\mu_0H^c$=2.4 T in (a), (b)and(c).}
  \label{fig:Fig5}
\end{figure}
To further investigate these novel PHE and PAMR, we define an effective field $\mu_0H^e=\mu_0H sin\theta$ along the $a$ axis which drives the carriers to move along its perpendicular direction by a Lorentz force within semi-classic model.  The $\rho_{xy}^P (\theta)$ and $\rho_{xx}^P (\theta)$ vs $\mu_0Hsin\theta$ are presented in Fig. 5. The weak kinks are observed in the butterfly-shape  $\rho_{xy}^P (\theta)$ curves at an effective critical field of $\mu_0H^c$=2.4 T. More clear features are observed in $\rho_{xx}^P (\theta)$ curves in Fig. 5(b). The sharp drops at various angle from Fig. 4(e) are scaled together at the same critical effective field  separating $\rho_{xx}^P$ curves into two regions:(1) the high-resistivity part with $|{\mu_0H^e}|$ \textless \, $\mu_0H^c$ and (2) the low-resistivity part with $|{\mu_0H^e}| \geq \mu_0H^c$. These behaviors are consistent with results acquired from the field dependent resistivity $\rho_{xy}^P (\mu_0 H)$ and magnetization $M(\mu_0 H)$ with $\mu_0 H // a$ exhibiting jumps or kinks around the magnetic flop transition from DS to FF states\cite{R. L. Dally}. The observed sudden jumps in $\rho_{xy}^P(\mu_0 H)$ indicate rotating the applied field can also drive same magnetic flop transition from DS to FF states which hosts the different scattering of carriers and changes the Zeeman gap leading to the different transport resistivity.

\begin{figure}
  \centering
  \includegraphics[width=3 in]{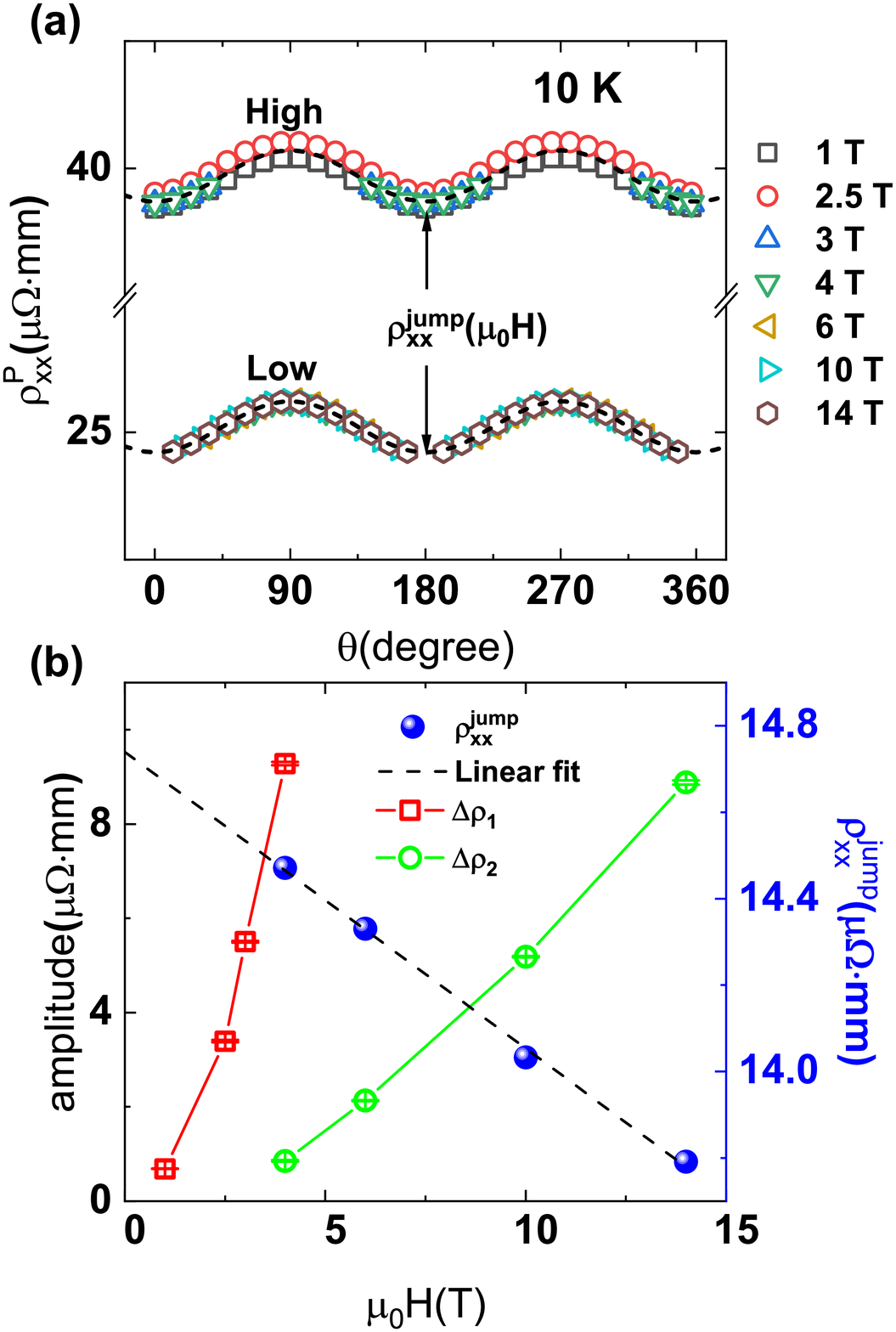}
  \caption{(a) The fitted parts of PAMR at 10 K with $\mu_0 H$=1 T, 2.5 T, 3 T, 4 T 6 T, 10 T, 14 T are subtracted from [$\rho_{0}^P -\rho_{xx}^{jump}(\mu_0 H)(|\mu_0H^e|\geq \mu_0H^c)$]  and divided by $\Delta \rho_n[n=1(|\mu_0H^e|\leq \mu_0H^c), n=2(|\mu_0H^e|\geq \mu_0H^c)$] to normalize the amplitude and add $\rho_{0}^P -\rho_{xx}^{jump}(\mu_0 H)(|\mu_0H^e|\geq \mu_0H^c)$ to clearly show the two stages. (b) $\Delta \rho_1$(red line), $\Delta \rho_2$(green line) and $\rho_{xx}^{jump}(\mu_0 H)$(blue symbol) as a function of $\mu_0 H$ at 10 K, black dash line is the fitting curve of $\rho_{\|}^P$.}
  \label{fig:Fig6}
\end{figure}
To describe the PAMR in presence of transitions, we first separate $\rho_{xx}^P (\theta)$ into the high- and low-resistivity groups. By using the relations $\rho_{xx}\sim sin2\theta$, $\rho_{xx}^P (\theta)$ for each group can be scaled together indicating the PAMR in each magnetic state following the experiential formula (3) independently as shown in Fig. 6(a). The amplitudes of oscillatory $\rho_{xx}^p (\theta)$ for  high- and low-resistivity groups exhibit separated field dependent behaviors. Usually in a magnetic system, PAMR is originated from the interaction of localized moments and spins of conduction electrons. In a single magnetic state, the interaction of magnetic moment and spin-orbit coupling varies little for changing $\mu_0 H$ leading to $\rho_{xx}^p (\theta)$ following the same field and angle dependence. When the system enters another magnetic state such as from the DS to FF state, the changed interaction of the magnetic moments and the spin-orbit coupling will host a separated field and angle dependence for PAMR. When rotating applied field $\mu_0 H$, the effective field $\mu_0 H^e$ is changing and drives the magnetic transition which makes the same effect as that by just changing the value of $\mu_0 H$ along the same direction of $\mu_0 H^e$. Moreover, it is found that the jump resistivity between DS and FF states $\rho_{xx}^{jump}(\mu_0H)$ ( defined as the height of the disconnection of $\rho_{xx}^P (\theta)$) exhibits linear field dependent variation instead of keeping a constant. That could be attributed to various effects on changing the carrier mobilities by rotating the $\mu_0 H$
and changing the value of $\mu_0 H^e$. The additional component parallel to the current for a rotated field  $\mu_0 H sin \theta$ can also change the scattering rate for carriers leading to a small increase for $\rho_{xx}^{jump}(\mu_0H)$. To describe PAMR with a magnetic transition, an extended experiential expression is developed as
\begin{equation}
\begin{aligned}
\rho_{xx}^P=&\rho_{0}^P -\rho_{xx}^{jump}(\mu_0 H)(|\mu_0H^e|\geq \mu_0H^c)\\ &+\Delta \rho_1 \cos ^2\theta(|\mu_0H^e|\leq \mu_0H^c)\\
&+\Delta \rho_2 \cos ^2 \theta(|\mu_0H^e|\geq \mu_0H^c)
\end{aligned}
\label{f4}
\end{equation}
where $\mu_0H^{e}$ is the effective field, $\rho_{xx}^{jump}$ is the  for high and low group $\Delta \rho_1$ and $\Delta \rho_2$ are the oscillatory amplitudes for different magnetic states. The data are fitted by this formula shown in Fig.4 (e).

\begin{figure}
  \centering
  \includegraphics[width=3.4in]{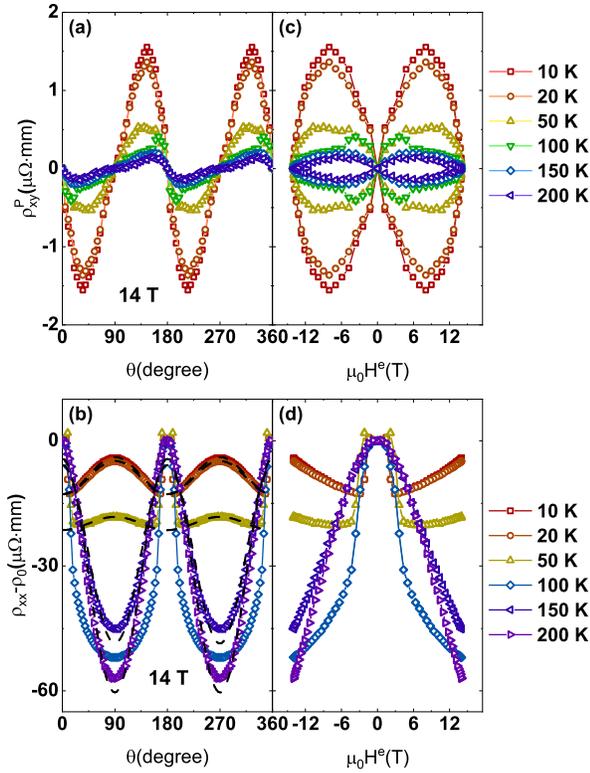}
  \caption{(a) The PHE $\rho_{xx}^P(\theta)$ at 10 K, 20 K, 50 K, 100 K, 150 K, 200 K with $\mu_0 H$=14 T. (b) Angular dependence of $\Delta \rho_{xx}^P(\theta)=\rho_{xx}^P(\theta)-\rho_{xx}^P(0^o)$ at 10 K, 20 K, 50 K, 75 K, 100 K, 150 K, 200 K with $\mu_0 H$=14 T.The black dash lines are fitting curve. (c) The PHE $\rho_{xx}^P(\mu_0 H^e)$ as a function of $\mu_0H^e=\mu_0H sin\theta$ at 10 K, 20 K, 50 K, 100 K, 150 K, 200 K with $\mu_0 H$=14 T. (d) $\Delta \rho_{xx}^P(\theta)$ as a function of $\mu_0H^e$ at 10 K, 20 K, 50 K, 75 K, 100 K, 150 K, 200 K with $\mu_0 H$=14 T.}
  \label{fig:Fig7}
\end{figure}
Fig. 7 shows the temperature dependence of $\rho_{xy}^P (\theta)$ and $\Delta \rho_{xx}^P(\theta)$ with $\mu_0 H$=14 T. The derivation of empirical laws  are observed in both $\rho_{xy}^P (\theta)$ and $\rho_{xx}^P (\theta)$ curves indicating the appearance of complicated magnetic states by changing $T$ in Figs. 7(a) and (b). By using the same scaling way, we analyze the planar transport properties evolving with changing  $T$ shown in Figs. 7(c) and (d). The $\mu_0 H^c$ relating to the magnetic flop shifts to lower field region and becomes invisible gradually with increasing $T$. By using the extended formula, most of the curves are fitted with little derivations. In fact, with changing $T$ the system undergoes multiple magnetic transitions even with the same $\mu_0 H$ resulting in various MR responses. It is observed with $\mu_0 H^e > \mu_0 H^c $ the MR exhibits positive response and evolves to negative response gradually with changing $T$. Thus at some temperatures, rotating $\mu_0 H$ will drive more than two magnetic states with much broad transitions which leads to fail to fit $\rho_{xx}^P (\theta)$ by a general rule. For TmMn$_6$Sn$_6$, Tm and Mn construct the system's magnetic structure together. The Tm$^{3+}$ ions
with 4$f$ electron states provide large magnetic moment (7.6 $\mu_B$)\cite{G. Venturini}, crystal field effect and strong magnetic interaction which hosts the more rich magnetic states leading to our observed novel transport properties. The further investigation for detailed magnetic structures at various temperatures and fields is necessary and expected to reveal these exotic states.

\section{summary}
We systematically  studied the transport properties of kagome magnet TmMn$_6$As$_6$. The observed prominent topological Hall effect in a wide temperature region suggests a non-zero spin chirality due to complicated non-collinear magnetic structure with strong fluctuations. In presence of the magnetic transitions, planar transport behavior (especial for PAMR) spanned over various magnetic states can be scaled within each magnetic state. Our results indicate the presence of multiple exotic magnetic states and provide the understandings of planar transport behaviors for general magnetic materials with crossovers between different magnetic states.

We thank Zhongbo Yan in Sun Yat-sen University for useful discussion. Work are supported by National Natural Science Foundation of Chin (NSFC) (Grants No.U213010013, 92165204, 11904414, 12174454),  Guangdong Basic and Applied Basic Research Foundation (Grant No. 2022A1515010035, 2021B1515120015), open research fund of Songshan Lake materials Laboratory 2021SLABFN11, OEMT-2021-PZ-02, National Key Research and Development Program of China (No. 2019YFA0705702).  and Physical Research Platform (PRP) in School of Physics, Sun Yat-sen University.

\end{document}